\documentclass[twocolumn,english,showpacs,floatfix]{revtex4}
\usepackage[dvips]{graphicx}
\usepackage{longtable}
\usepackage{amsmath,amssymb}
\usepackage{dcolumn}

\usepackage{babel}
\usepackage[latin1]{inputenc}
\usepackage{dcolumn}% Align table columns on decimal point
\begin{document}

%\preprint{}

\title{Quantum stress tensor for massive vector field \\
in the space-time of a cylindrical black hole}
%%%%%%%%%%%%%%%%%%%%%%%%%%%%%%%%%%%%%%%%%%%%%%%
%%%%%%%%%%%%%%%%%%%%%%%%%%%%%%%%%%%%%%%%%%%%%%%
\author{Owen Pavel Fern\'{a}ndez Piedra}
%%%%%%%%%%%%%%%%%%%%%%%%%%%%%%%%%%%%%%%%%%%%%%%
\email{opavel@ucf.edu.cu }
%%%%%%%%%%%%%%%%%%%%%%%%%%%%%%%%%%%%%%%%%%%%%%%
\affiliation{$^{1}$ Departamento de F\'{i}sica y Qu\'{i}mica, Universidad
de Cienfuegos, Carretera a Rodas, Cuatro Caminos, s/n. Cienfuegos,
Cuba,}
\affiliation{$^{2}$ Instituto de F\'{\i}sica, Universidade de S\~ao Paulo,
  CP 66318,
05315-970, S\~ao Paulo, Brazil}
%%%%%%%%%%%%%%%%%%%%%%%%%%%%%%%%%%%%%%%%%%%%%%%
%%%%%%%%%%%%%%%%%%%%%%%%%%%%%%%%%%%%%%%%%%%%%%%

\author{Jerzy Matyjasek}
\email{jurek@kft.umcs.lublin.pl}
\affiliation{Institute of Physics, Maria Curie-Sklodowska University
pl. Marii Curie Sklodowskiej 1, 20-031 Lublin, Poland.}

\begin{abstract}

 \noindent The components of the renormalized quantum Energy-Momentum tensor
for  a massive vector field coupled  to the gravitational field
configuration of a static Black-String  are analytically evaluated
using the Schwinger-DeWitt approximation. The general results are
employed to investigate the pointwise energy conditions for the
quantized matter field, and it is shown that they are violated at
some regions of the spacetime, in particular the horizon of the
black hole.
\end{abstract}

%\keywords{}

\maketitle

Quantum theory and General Relativity are two beautiful parts of
modern physics that, for more than a century have been developed in
such an extent that our knowledge of the universe at short and long
scales has increased as never before in the human history. With the
help of the quantum theory we can explain micro-world phenomena. On
the other hand, the General Theory of Relativity allows us a deep
understanding of the large scale structure of the universe. This two
major achievements in theoretical physics in the 20th century, are
still, nearly 100 years later, going separated ways. There is not
yet such a thing as a theory of quantum gravity, but in their quest
for the TOE (Theory of Everything), the physicists try to bring them
together. Quantum gravitation is a tool that would be very important
to describe, among other things, the creation of the universe and
its later development.

One of the approaches developed to consider quantum effects in
gravitation, called Semiclassical gravity, considers the quantum
dynamics of fields in a gravitational background, which at this
level of description is considered as a classical external field. In
the absence of a full theory of quantum gravity, semiclasical
gravity is a well established physical theory that help us to know
what are the expected behavior of gravitational system under the
influence of the interaction between it and matter fields that obeys
the laws of quantum theory.

In this approximate theory, fundamental information about the
quantum matter fields is contained in the renormalized quantum
stress-energy tensor \(\langle T_{\mu}^{\nu}\rangle_{ren}\), that
can, in principle, be constructed using a variety of mathematical
techniques, including analytical, semianalytical and numerical ones,
see
\cite{dowker,browncassidy,allenfolacci,howardcandelas,candelas,fawcet,jensen12,AHS,frolov,DeWitt,avramidi,barvinsky,matyjasek,
matyjasek1,owencabo1,owencabo2} and references therein.

For the important case of massive fields, one of the developed
approaches for determining \(\langle T_{\mu}^{\nu}\rangle_{ren}\) is
based in the calculation of the renormalized quantum effective
action for the quantized matter field, using the Schwinger-DeWitt
proper time technique to give an expansion of the effective action
in terms of the field inverse square mass. This is the celebrated
Schwinger-DeWitt expansion, in which the first three terms
renormalize the bare gravitational and cosmological constant, and
adds some higher order terms to the Einstein gravitational action.
The next order term, proportional to $m^{-2}$, where $m$ is the
mass of the field, gives us the one-loop effective action $W_{ren}$
for the matter field
\cite{frolov,DeWitt,avramidi,barvinsky,matyjasek,
matyjasek1,owencabo1,owencabo2}.

By functional differentiation of the one-loop effective action, we
can obtain the desired quantum stress tensor using the standard
formula

\begin{equation}\label{}
    \langle T_{\mu\nu}\rangle_{ren}=\frac{2}{\sqrt{\ -g}}\frac{\delta W_{ren}}{\delta g^{\mu\nu}}
\end{equation}

The above method has been applied to a number of space-times of
interest, including Schwarzschild \cite{frolov, AHS},
Reisner-Nordstrom \cite{AHS,matyjasek}, charge dilatonic black holes
and nonlinear electrically charged black holes in four dimensions
\cite{matyjasek1}. Also, in two recent papers we developed the
Schwinger-DeWitt technique for the calculation of the renormalized
stress energy tensor of massive scalar and spinor fields up to one
loop order in the spacetime of static black strings
\cite{owencabo1,owencabo2}. For this interesting system, the
problems of investigate the renormalized stress tensor components
for conformally coupled massless scalar fields were studied by
DeBenedictis in \cite{debenedictis1,debenedictis2}, who used the
obtained \(\langle T_{\mu}^{\nu}\rangle_{ren}\) for the calculation
of gravitational backreaction of the quantum field. In this work we
complete the series of papers \cite{owencabo1,owencabo2} dedicated
to the calculation of \(\langle T_{\mu}^{\nu}\rangle_{ren}\) for
massive fields in the static black string background, determining
the components of this tensor for the case of a massive vector
field. We also investigate the fulfilment of the pointwise energy
conditions for the quantized field in this gravitational background.

The corresponding metric element for the static black string
spacetime is
\begin{widetext}
\begin{equation}\label{}
    ds^{2}=-(\alpha^{2}\rho^{2}-\frac{4M}{\alpha\rho})dt^{2}+
    \frac{1}{(\alpha^{2}\rho^{2}-\frac{4M}{\alpha\rho})}d\rho^{2}+
    \rho^{2}d\varphi^{2}+\alpha^{2}\rho^{2}dz^{2}.  \label{ds2simp}
\end{equation}
\end{widetext}
where \(M\) is the mass per unit length of the string. As we can see
from (\ref{ds2simp}),  the considered metric has an event horizon
located at \(\rho_{+}=\frac{\sqrt[3]{4M}}{\alpha}\)and the only true
singularity is a polynomial one at the origin.

The action for a single massive vector field \(A_{\mu}\) with mass
\(m\) in some generic curved spacetime in four dimensions is
\begin{equation}\label{}
    S=-\int d^{4}x\sqrt{-g}\left(\frac{1}{4}F_{\mu\nu}F^{\mu\nu}+\frac{1}{2}m^{2}A_{\mu}A^{\mu}\right)\label{fieldaction}
\end{equation}
The equation of motion for the field have the form
\begin{equation}\label{}
    \hat{D}^{\mu}_{\nu}\left(\nabla\right)A_{\mu}=0 \label{fieldeqn1}
\end{equation}
where the second order operator
\(\hat{D}^{\mu}_{\nu}\left(\nabla\right)\) is given by
\begin{equation}\label{}
    \hat{D}^{\mu}_{\nu}\left(\nabla\right)=\delta_{\nu}^{\mu}\Box-\nabla_{\nu}\nabla^{\mu}-R_{\nu}^{\mu}-m^{2}\delta_{\nu}^{\mu}\label{nonminimal}
\end{equation}
where \(\Box\,=\,g^{\mu\nu}\nabla_{\mu}\nabla_{\nu}\) is the
covariant D'Alembert operator, \(\nabla_{\mu}\) is the covariant
derivative.

The usual formalism of Quantum Field Theory give an expression for
the effective action of the quantum field \(A_{\beta}\) as
perturbation expansion in the number of loops:
\begin{equation}\label{}
    \Gamma\left(A_{\beta}\right)=S\left(A_{\beta}\right)+\sum_{k\geq1}\Gamma_{(k)}\left(A_{\beta}\right)
\end{equation}
where \(S\left(A_{\beta}\right)\) is the classical action of the
free field. The one loop contribution of the field \(A_{\beta}\) to
the effective action is expressed in terms of the operator
(\ref{nonminimal}) as:
\begin{equation}\label{}
    \Gamma_{(1)}=\frac{i}{2}\ln\left(\mathfrak{Det}\hat{D}\right)
\end{equation}
where \(\mathfrak{Det}\hat{F}=\exp(\mathbb{T}\mathrm{r}\ln\hat{F})\)
is the functional Berezin superdeterminant \cite{avramidi} of the
operator \(\hat{F}\), and \(\mathbb{T}\mathrm{r}
\hat{F}=\left(-1\right)^{i}F^{i}_{i}=\int
d^{4}x\left(-1\right)^{A}{F}^{A}_{A}(x)\) is the functional
supertrace \cite{avramidi}. If the Compton's wavelength of the field
is less than the characteristic radius of spacetime curvature
\cite{frolov,barvinsky,DeWitt,avramidi,matyjasek,matyjasek1,owencabo1,owencabo2}, we can develope an
expansion of the above effective action in powers of the inverse
square mass of the field. This approximation is known as the
Schwinger-DeWitt one, and before applying this approach to the
particular problem considered in this work we make the following
remarks. In the first place, we mention that the Schwinger-DeWitt
technique is directly applicable to "minimal" second order
differential operators that have the general form:
\begin{equation}\label{}
     \hat{K}^{\mu}_{\nu}\left(\nabla\right)=\delta_{\nu}^{\mu}\Box-m^{2}\delta_{\nu}^{\mu}+Q^{\mu}_{\nu} \label{minimal}
\end{equation}
where \(Q^{\mu}_{\nu}(x)\) is some arbitrary matrix playing the role
of the potential.

As we can see, because of the presence of the nondiagonal term in
(\ref{nonminimal}) it becomes a nonminimal operator, and this fact
is an obstacle to applying the Schwinger-DeWitt technique. By
fortune we can put (\ref{nonminimal}) as function of some minimal
operators, if we note that it satisfies the identity
\(\hat{D}^{\mu}_{\nu}\left(\nabla\right)\left(m^{2}\delta_{\nu}^{\mu}-\nabla_{\nu}\nabla^{\mu}\right)=m^{2}\left(\delta_{\nu}^{\mu}\Box-R_{\nu}^{\mu}-m^{2}\delta_{\nu}^{\mu}\right)\).
\begin{widetext}
Then the one loop effective action for the nonminimal operator
(\ref{nonminimal}) omitting an inessential constant can be written
as
\begin{equation}\label{}
   \frac{
   i}{2}\mathbb{T}\mathrm{r}\ln\hat{D}^{\mu}_{\nu}\left(\nabla\right)=\frac{
   i}{2}\mathbb{T}\mathrm{r}\left(\delta_{\nu}^{\mu}\Box-R_{\nu}^{\mu}-m^{2}\delta_{\nu}^{\mu}\right)-\frac{
   i}{2}\mathbb{T}\mathrm{r}\left(m^{2}\delta_{\nu}^{\mu}-\nabla_{\nu}\nabla^{\mu}\right)\label{split}
\end{equation}
We can see in (\ref{split}) that the first term is the effective
action of a minimal second order operator
\(K^{\mu}_{\nu}\left(\nabla\right)\) with potential
\(-R_{\nu}^{\mu}\). The second term can be transformed as
\(\mathbb{T}\mathrm{r}\left[\frac{1}{m^{2}}\nabla^{\mu}\nabla_{\nu}\right]^{n}=\mathbb{T}\mathrm{r}\left[\frac{1}{m^{2}}\nabla^{\mu}\Box^{n-1}\nabla_{\nu}\right]=\mathbb{T}\mathrm{r}\left[\frac{1}{m^{2}}\Box\right]^{n}\)
and \begin{equation}\label{}
    \frac{
   i}{2}\mathbb{T}\mathrm{r}\left(m^{2}\delta_{\nu}^{\mu}-\nabla_{\nu}\nabla^{\mu}\right)=\frac{
   i}{2}\mathbb{T}\mathrm{r}\left(m^{2}-\Box\right)
\end{equation}
\end{widetext}
Then, the effective action for the massive vector field is equal to
the effective action of the minimal second order operator
\(K^{\mu}_{\nu}\left(\nabla\right)\) minus the effective action of a
minimal operator \(S^{\mu}_{\nu}\left(\nabla\right)\) corresponding
to a massive scalar field minimally coupled to gravity.

Now using the Schwinger-DeWitt representation for the Green´s
function of the minimal operators, we can obtain for the
renormalized one loop effective action of the quantum massive vector
field the expression \(\Gamma_{(1) ren}\,=\,\int  d^{4}x
\sqrt{-g}\,\mathfrak{L}_{ren}\) where the renormalized effective
Lagrangian reads:
\begin{equation}
\mathfrak{L}_{ren}\,=\,{1\over 2(4\pi)^{2}\,}
\sum_{k=3}^{\infty}{\left(\,\mathbb{T}\mathrm{r}
\,a^{(1)}_{k}(x,x)-\,\mathbb{T}\mathrm{r}
\,a^{(0)}_{k}(x,x)\right)\over
k(k-1)(k-2)m^{2(k-2)}}\label{renlagrangian},
\end{equation}
The quantities \([a^{(1)}_{k}]= \,a^{(1)}_{k}(x,x')\) and \([a^{(0)}_{k}]=
\,a^{(0)}_{k}(x,x')\), whose coincidence limit appears under the
supertrace operation in (\ref{renlagrangian}) are the HMDS
coefficients for the minimal operators
\(K^{\mu}_{\nu}\left(\nabla\right)\) and
\(S^{\mu}_{\nu}\left(\nabla\right)\) respectively. As usual, the
first three coefficients of the DeWitt-Schwinger expansion,
$a_{0},\,a_{1},\,{\rm and}\,a_{2}, $ contribute to the divergent
part of the action and can be absorbed in the classical
gravitational action by renormalization of the bare gravitational
and cosmological constants.

Restricting ourselves here to the terms proportional to $m^{-2},$
using integration by parts and the elementary properties of the
Riemann tensor
\cite{avramidi,matyjasek,matyjasek1,owencabo1,owencabo2}, we obtain
for the renormalized effective lagrangian in the case of the massive
vector field considered in this work
\begin{widetext}
\begin{eqnarray}
 \nonumber
  \mathfrak{L}_{ren}&=&{1\over 192 \pi^{2} m^{2}} \left[{9\over 28} R_{\mu \nu} \Box R^{\mu \nu}- \frac{27}{280} R
 \Box R-{5\over 72} R^{3}+{31\over 60} R R_{\mu \nu } R^{\mu \nu}-{52\over 63} R^{\mu}_{\nu} R^{\nu}_{\gamma} R^{\gamma}_{\mu}-{19\over 105} R^{\mu \nu}
 R_{\gamma \varrho} R^{\gamma ~ \varrho}_{~ \mu ~ \nu}
\right. \nonumber \\ &&\left. \,+\ {61\over 140} R_{\mu \nu}
R^{\mu}_{~ \sigma \gamma \varrho} R^{\nu \sigma \gamma \varrho}
-{1\over 10}R R_{\mu \nu \gamma \varrho} R^{\mu \nu \gamma
\varrho}-{67\over 2520} {R_{\gamma \varrho}}^{\mu \nu} {R_{\mu
\nu}}^{\sigma \tau} {R_{ \sigma \tau}}^{\gamma \varrho}\,+\,{1\over
18} R^{\gamma ~ \varrho}_{~ \mu ~ \nu} R^{\mu ~ \nu}_{~ \sigma ~
\tau} R^{\sigma ~ \tau}_{~ \gamma ~
\varrho}\right]\label{renlagrangian1}
\end{eqnarray}
\end{widetext}
As we can see, this final expression of the one loop effective for
the massive vector field only differ from that of the massive scalar
and spinor fields in the numerical coefficients in front of the
purely geometric terms. For \(\langle T_{\mu\nu}\rangle_{ren}\) we
obtain a very cumbersome expression that, as in the case of
(\ref{renlagrangian1}), is different from that obtained for scalar
and spinor fields only in the numerical coefficients that appears in
front of the purely geometrical terms. For this reason we not put
this very long expression for the stress tensor here and refers the
readers to our previous papers \cite{matyjasek,matyjasek1} and
\cite{owencabo1,owencabo2}.

 It is interesting to mention that in a beautiful paper
 Dec\'{a}ninis and Folacci \cite{Folacci} have presented irreducible
expressions for the metric variations of the gravitational action
terms constructed from the 17 curvature invariants of order six in
derivatives of the metric tensor i.e. from the geometrical terms
appearing in the diagonal coefficient $a_{3}(x,x)$ of the Schwinger-DeWitt
approximation, thus providing us with a general method to reduce the
inevitable differences in the final expresions obtained for this quantities, due to the different simplification and
canonization schemes chosen. %

From the general form of the geometric terms conforming the general
expresion for the constructed \(\langle
T_{\mu}^{\nu}\rangle_{ren}\), we see that it is covariantly
conserved, thus indicating that it is a god candidate for the
expected exact one in our large mass approximation.

\begin{figure}
  \includegraphics[width=8cm]{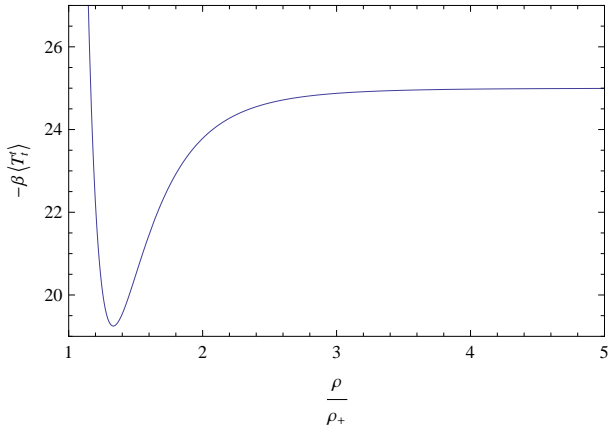}\\
  \caption{Radial dependence of the rescaled component of the energy density \(\varrho=-\left\langle T_{t}^{t}\right\rangle\) of the quantum massive vector field in the geometry of a static black string. The coefficient \(\beta=3360\pi^{2}m^{2}\alpha^{-6}\).}
\label{figure1}
\end{figure}

\begin{figure}
  \includegraphics[width=8cm]{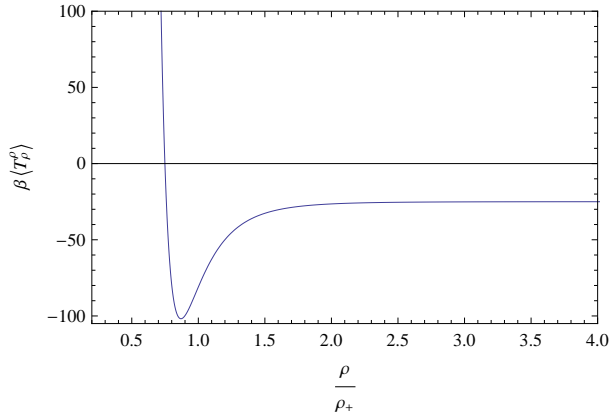}\\
  \caption{Radial dependence of the rescaled component \(\left\langle T_{\rho}^{\rho}\right\rangle\) of the quantum massive vector field in the geometry of a static black string. The coefficient \(\beta=3360\pi^{2}m^{2}\alpha^{-6}\).}
\label{figure2}
\end{figure}

\begin{figure}
  \includegraphics[width=8cm]{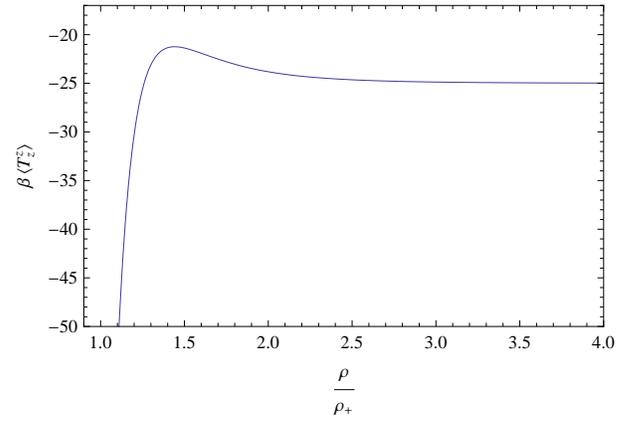}\\
  \caption{Radial dependence of the rescaled component \(\left\langle T_{z}^{z}\right\rangle\) of the quantum massive vector field in the geometry of a static black string. The coefficient \(\beta=3360\pi^{2}m^{2}\alpha^{-6}\).}
\label{figure3}
\end{figure}

After a direct calculation, we obtain for \(\langle
T_{\mu}^{\nu}\rangle_{ren}\) in the space-time of a static
cylindrical black hole metric
\begin{equation}\label{}
    \left\langle T_{\mu}^{\ \mu}(y)\right\rangle_{ren}=\frac{1}{3360 \pi^{2}m^{2}\alpha^{-6}}\left(a_{\mu}
    +\frac{\Lambda_{\mu}}{y^{6}}+\frac{\Omega_{\mu}}{y^{9}}\right),\label{set}
\end{equation}
where we have defined the variable \(y=\frac{\rho}{\rho_{+}}\) and
due to the cylindrical symmetry we have \(\left\langle T_{z}^{\
z}\right\rangle_{ren}=\langle T_{\varphi}^{\
\varphi}\rangle_{ren}\). The numerical coefficients are given in
Table I for each index \(\mu\). The dependence of the components of
\(\left\langle T_{\mu}^{\ \mu}\right\rangle_{ren}\) with \(y\) is
displayed in figures (\ref{figure1}) to (\ref{figure3}).

\begin{table}[htb!]\label{Tabla1}
  \begin{center}
         % tamanho da fonte
   \setlength{\arrayrulewidth}{2\arrayrulewidth}  % espessura da  linha
   \setlength{\belowcaptionskip}{1pt}  % espaÃ§o entre caption e tabela
  \begin{tabular}{|c|c|c|c|}
    \hline
     \ \ \ \ $\mu$  \ \ \ \  &   \ \ \ \ \(a_{\mu}\) \ \ \ \  &  \ \ \ \ \(\Lambda_{\mu}\) \ \ \ \   &  \ \ \ \ \ \(\Omega_{\mu}\) \ \ \ \ \\
     \hline
    t & -25 & $387/4$& - 611/4 \\
    \hline
    \(\rho\) & -25& - 399/4 & 175/4 \\
    \hline
    z & -25 & 405/4& - 809/4 \\
    \hline
  \end{tabular}
   \caption{\it Numerical coefficients in the general expresion for the quantum stress tensor of massive vector field in the spacetime of cylindrical black hole.}
   \end{center}
\end{table}

If we consider the general expression (\ref{set}) at the horizon of
the cylindrical black hole, i.e, at \(y=1\), we easily found that
the energy density \(\varepsilon=-\left\langle T_{t}^{\
t}\right\rangle_{ren}\) for the quantum massive vector field is
positive, in contrast with the results found in previous work for
the scalar and spinor fields \cite{owencabo1,owencabo2}.

Inspection of figure (\ref{figure1}) shows that the energy density
is positive everywhere. The principal pressures
\(p_{1}=-\tau=\left\langle T_{\rho}^{\ \rho}\right\rangle_{ren}\)
and \(p_{2}=p_{3}=p=\left\langle T_{z}^{\ z}\right\rangle_{ren}\)
are negative at the horizon. Figures (\ref{figure2}) and
(\ref{figure3}) indicates that the radial pressure is negative in
the region outside the horizon and that the other pressures are
negative everywhere. At the event horizon \(\varrho-\tau=0\),
\(\varrho+p<0\) and \(\varrho-\tau+2p<0\). Also we have
\(p<-\varrho<\varrho\). The second of the above relations indicates
that the null energy condition (NEC) is violated at the event
horizon of the static black string. For the weak energy condition
(WEC) be satisfied, we need that the energy density be positive, as
is indeed the case at the horizon, but we require that the NEC be
satisfied. Then, in our case, also the weak energy conditions is
violated. If the NEC is satisfied and the sum of the principal
pressures and the energy density of the field is positive, then the
strong energy condition (SEC) is valid. The dominant energy
condition (DEC) requires \(-\varrho\leq p_{j}\leq\varrho\). As we
can see from the above relations between the energy density and the
principal pressures at the horizon of the black string, the massive
vector field also violates the SEC and DEC.

The results of this work are expected to be employed to investigate
the back-reaction of the quantum scalar field, on the Black String
metric. For this purpose, the Einstein equations for the metric
should be solved after including in them the calculated stress
tensor for the Black String solution. Our results of the
implementation of this programm will be published in the future.

\end{document}